\documentstyle[newarcrc,fleqn,psfig]{article}

\def\msun{$M_{\odot}$}
\def\mdot{$\dot M$}

\def\etal{{\it et al.}}

\def\grad{$^\circ$}

\def\asec{\ifmmode ^{\prime\prime}\else$^{\prime\prime}$\fi}
\def\amin{\ifmmode ^{\prime}\else$^{\prime}$\fi}
\def\degs{\ifmmode ^{\circ}\else$^{\circ}$\fi}
\newbox\grsign \setbox\grsign=\hbox{$>$}
\newdimen\grdimen \grdimen=\ht\grsign
\newbox\laxbox \newbox\gaxbox
\setbox\gaxbox=\hbox{\raise.5ex\hbox{$>$}\llap
     {\lower.5ex\hbox{$\sim$}}}\ht1=\grdimen\dp1=0pt
\setbox\laxbox=\hbox{\raise.5ex\hbox{$<$}\llap
     {\lower.5ex\hbox{$\sim$}}}\ht2=\grdimen\dp2=0pt

\def\grs{GRS 1915+105}

\title{RXTE Spectroscopy of GRS 1915+105}

\author{Jochen Greiner\address{Astrophysikalisches Institut Potsdam, 
   14482 Potsdam, Germany},
   Ed H. Morgan$^{\rm b}$, Ron A. Remillard\address{Center for Space Research, 
          MIT, Cambridge, MA 02139, USA }}

\begin{document}
\maketitle

\begin{abstract}
The galactic superluminal motion source GRS 1915+105 has been extensively
observed with the RXTE satellite over the last two years.
More than 250 RXTE pointings have been performed until mid-May 1998 with
more than 1.5 Msec exposure time on roughly a weekly basis.
Here we report on first results of our spectral analysis of a major part of
these pointed RXTE observations. We establish the existence of at least 
5 spectral components and present the changes of these components over the last
two years.
\end{abstract}

\section{Introduction}

GRS 1915+105 was discovered in 1992  with the WATCH/{\it Granat} 
(Castro-Tirado \etal\ 1992). A comparison of the BATSE ($>25$ keV) fluxes with 
ROSAT (1--2.4 keV) fluxes has shown that 
GRS 1915+105 has been active all the time since 1992, even during times 
of BATSE non-detections (Greiner \etal\ 1997). This has become even more
evident with the daily coverage provided by the all-sky monitor (ASM) of
the RXTE satellite since early 1996 (see top of Fig.\,\ref{fitpar}).
A variable radio source was found with the VLA (Mirabel \etal\ 1993)
inside the $\pm$10\asec\ X-ray error circle (Greiner 1993), which later was 
discovered to eject radio blobs travelling at apparently 
superluminal speed (Mirabel \& Rodriguez 1994) making GRS 1915+105 the first
superluminal source in the Galaxy. Until then, apparent superluminal motion
was only observed in AGN, the central engines
of which are generally believed to be massive black holes. This similarity
as well as that to the second superluminal source in our Galaxy, GRO J1655--40,
for which a dynamical mass  of 7 \msun\ was estimated (Orosz \& Bailyn 1997) 
suggests that GRS 1915+105 harbors a stellar-sized black hole.
The series of RXTE pointed observations initiated in April 1996 revealed
dramatic intensity variations (Greiner \etal\ 1996) on time scales from 10\,s 
to hours. The nature of these astonishing X-ray instabilities is
currently a mystery though attempts have been made to both interprete these as
accretion disk instabilities leading to an infall of parts of the 
inner accretion disk (Greiner \etal\ 1996, Belloni \etal\ 1997) and relate
them to radio flares and jet formation (Greiner \etal\ 1996, 
Pooley \& Fender 1997, Eikenberry \etal\ 1998, Mirabel \etal\ 1998).

The X-ray spectrum as seen with  ROSAT (Greiner 1993) and ASCA (Nagase \etal\
1994) is strongly absorbed (N$_{\rm H}\approx 5\times10^{22}$ cm$^{-2}$) 
consistent with the location in the galactic plane at 12.5 kpc distance
(Mirabel \& Rodriguez 1994). Complex emission/absorption features in the
7--8 keV range have been seen with ASCA at two occasions (Ebisawa 1997).

\section{Spectral analysis}

We have started a comprehensive spectral investigation using PCA as well
as HEXTE data of well-defined
time stretches which are selected according to their different shapes
in the lightcurve. We used the ``Standard 2'' data having 16 sec time
resolution, and used PCUs 0, 3 and 4 separately in the fits to not degrade
the spectral resolution. We have always fit combined PCA and HEXTE data.
We have ignored in the fitting PCA data below 2 keV and above 25 keV due to 
uncertainties in the response matrix
(see http:/$\!$/lheawww.gsfc.nasa.gov/users/keith/user\_comm/sept97/pca.html 
for a detailed description) and HEXTE data below 30 keV due to response
uncertainties and above 170 keV due to count rate statistics.
We have added a 1\% systematic error to account for the remaining
PCA response uncertainties.
While applying the identical model to both data sets, we let the relative
normalization between the PCA and HEXTE data float as a free parameter.
No deadtime correction has been applied which may result in an underestimate
of the fluxes in the PCA of about 5\% during the very high flux states.

The spectra are complex and rapidly variable. 
Single component spectra like pure power law, brems\-strahlung, 
synchrotron or comptonization models do not fit these spectra. 
In general, the spectra are composed of at least five components:
\begin{itemize}
\vspace{-0.28cm}\item {\bf Soft component:} 
The soft component is a transient and most probably thermal component
below $\sim$15 keV. It can formally be described equally well by either
an exponential cut-off power law (with cut-off energy between 5--7 keV),
a bremsstrahlung model with typical temperatures in the range 3--5 keV, 
or a multicolor disk blackbody model. The bremsstrahlung model description, 
however, is physically not possible because the observed luminosity
requires a size of the emission region for which the light crossing time 
would be at least a factor of 10--30 larger than the observed variability 
time scale. We therefore adopt the multicolor disk blackbody model
description in the following. The typical effective temperature of the 
disk blackbody is 1--2 keV. Note that in its standard form (like DISK
within XSPEC) the best fit accretion rate is 
always highly super-Eddington, and the implied mass of the compact object
below 1 \msun\ due to the high temperature.
\vspace{-0.28cm}\item {\bf Hard component:}
The hard component is best described by a power law model extending up to 200
keV, the upper bound of the HEXTE energy range. This component is always
present, but varies in strength and power law photon index (between --2.0 to
--4.8). 
\vspace{-0.28cm}\item {\bf Compton reflection component:}
There is an nearly always visible additional component comprising of
excess emission in the 10--20 keV range which we interpret as
Compton reflection. 
\vspace{-0.28cm}\item {\bf Iron complex:}
We observe  strong excess emission in the 6--8 keV region. The peak of
this excess emission  varies between 6.5 and 7.5 keV. Due to the PCA
energy resolution of about 0.9 keV in this range a more detailed analysis,
such as revealing various components, is not possible.
\vspace{-0.28cm}\item {\bf Correction for dust scattering:}
\grs\ is located behind $\sim$5$\times$10$^{22}$ cm$^{-2}$ of absorbing
material (Greiner \etal\ 1994), and thus exhibits a strong dust scattering 
component. Based on ROSAT PSPC observations the fraction of scattered light
at 1 keV is about 30\% and the halo size is larger than the field of view
of the PSPC (55\amin\ radius). We note that due to the strong X-ray variability
of \grs\ 
this value is not a constant since the scattered light travels increasingly 
longer as one moves to increasing angular distance from the line of sight.
Since the collimated field of view of the PCA is smaller than the dust
scattering halo size of \grs\ one looses X-ray source photons which are
scattered off the line of sight and arrive at Earth at an angle larger than the
PCA field of view radius. The result of correcting for this effect (with the
DUST model in XSPEC) is a reduction of the best-fit absorbing column
by as much as a factor of 1.5--2. The best-fit absorbing column then
ranges between 5.3--5.7$\times$10$^{22}$ cm$^{-2}$ in our fits, 
consistent with the ROSAT, ASCA and radio measurements of the extinction
towards \grs.
\end{itemize}\vspace{-0.28cm}

\begin{figure}
      \vbox{\psfig{figure=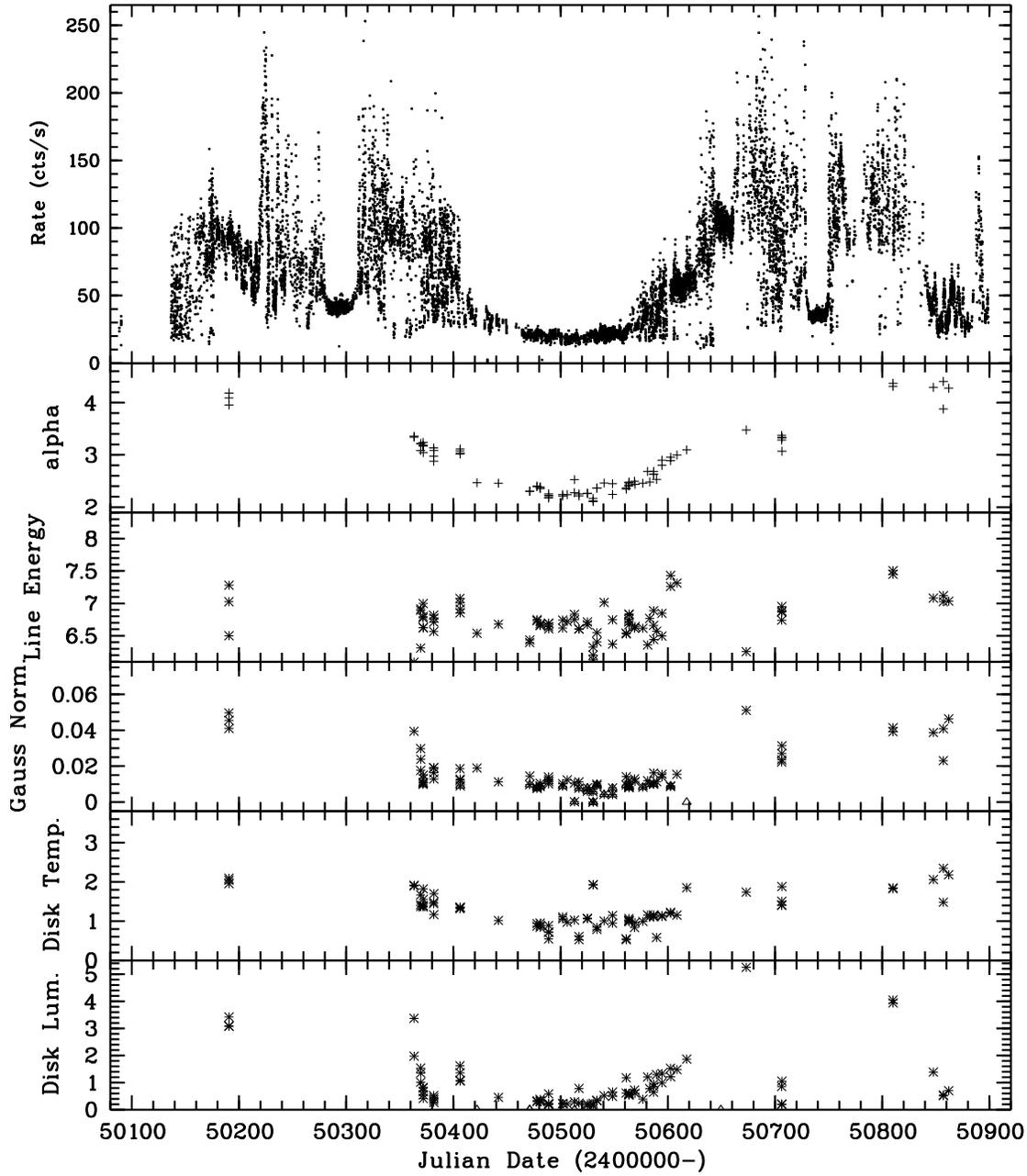,width=14.7cm,%
          bbllx=2.2cm,bblly=4.9cm,bburx=19.1cm,bbury=24.6cm,clip=}}\par
      \vspace{-0.5cm}
      \caption[fitpar]{Top panel: RXTE ASM lightcurve of GRS 1915+105
        since February 1996 in the 2--12 keV range (1 Crab corresponds
        to about 73 cts/s). Lower panels: Variation of fit parameters
        with time for a spectral model consisting of the sum
        of a multicolor disk blackbody, a power law with a reflection
        component, a Gaussian emission line and correction for the dust
        scattering. The parameters plotted are
        from top to bottom: power law photon index alpha, central energy
        (in keV) and normalisation (intensity)  of the  Gaussian emission 
        line (in  ph/cm$^2$/s), 
        and the temperature (1.2$\times$ [\mdot/M]$^{1/4}$ keV) and
        luminosity (\mdot\ $\times$ M $\times$ normalisation) of 
        the multicolor disk blackbody. 
         }
         \label{fitpar}
    \end{figure}

\noindent
We have fit the spectra of more than 50 observations with a combination
of these five components, and show the temporal variation of some basic 
best-fit parameters in Fig. \ref{fitpar}.

The RXTE observation of 1997 October 31 shows \grs\ in a state of
wild oscillations with the largest peak fluxes we have observed so far.
The count rate in all PCA units reaches 63\,040 cts/s during these peaks which
have a typical duration of 10--12 sec, and which have complex peak structure
(flat-top, round, double-peaked). The determination of the total X-ray flux
depends on the lower bound of the power law component. The best-fit --3.2 power
law component contributes 2.3$\times$10$^{-7}$ erg/cm$^2$/s
(3.7$\times$10$^{-6}$ erg/cm$^2$/s) in the 1--200 keV (0.1--200 keV) range.
The bolometric flux of the $\sim$2 keV disk blackbody component is 
1.6$\times$10$^{-7}$ erg/cm$^2$/s.
Thus, integrating over the disk blackbody component plus the power law
between the usually adopted lower bound of 0.1 keV up to 200 keV 
results in an unabsorbed luminosity 
of 6.5$\times$10$^{40}$ (D/12.5 kpc)$^2$ erg/s (or
6.7$\times$10$^{39}$ (D/12.5 kpc)$^2$ erg/s for a lower bound of 1 keV).
The average luminosity over several hours is about one third of this.

\section{Discussion and Summary}

The slope of the power law component shows a clear variation which is
correlated to the intensity states: it is flattest during the low-hard
states (Morgan \etal\ 1997), and much steeper during the high and flaring
states. In general, the power law component extending to above 100 keV is 
assumed to be produced by
Comptonization of the soft disk X-ray photons by the hot electrons 
suspected to exist in the disk ``corona''. Given the high luminosity of 
this power law component (see above), Comptonization must be very efficient.
However, the more efficient Comptonization is the flatter should the resulting
power law slope be, inconsistent with our observations. We therefore conclude
that other emission mechanisms should be considered, e.g. synchrotron emission 
as one of the possible alternatives.

Using an inclination of 70\grad\ and the
Compton reflection models in the XSPEC package which consistently
include an iron absorption edge we find that
(i) the reflection fraction is rather low and that
(ii) an ionized reflector seems to be required since the ionization
parameter (in the PEXRIV model) always peggs the upper limit.
We note, however, that at some occasions the Gaussian line component
goes to the iron absorption edge to seemingly compensate for a smaller
or even non-existing edge structure.
The inclusion of the reflection component in the fit has the general effect
of resulting in a steeper power law slope, i.e. the reflected emission
component in the 10--30 keV range does not compromise the actual slope
in the HEXTE energy range up to 170 keV. In addition, it also reduces
the accretion rate, and thus temperature, of the disk blackbody component
by about 20\%.

\vspace{0.3cm}\noindent {\it Acknowledgements:}
JG is supported by the German Bundesmi\-ni\-sterium f\"ur Bildung, 
Wissenschaft, Forschung und Technologie (BMBF/DLR) under contract No. 
FKZ 50 QQ 9602 3.


\begin{thebibliography}{Davidson \& Humphreys, 1997}

\bibitem[]{bell97} Belloni T., Mendez M., King A.R., van der Klis M.,
  van Paradijs J., 1997, ApJ 479, L145

\bibitem[]{ct92} Castro-Tirado A.J., Brandt S., Lund N., 1992, IAU Circ. 5590


\bibitem[]{e97} Ebisawa K., 1997, 
in Proc. of ``X-ray imaging and spectroscopy of 
cosmic hot plasmas'', Univ. Acad. Press, Tokyo, p. 427

\bibitem[]{emm98} Eikenberry S.S., Matthews K., Morgan E.H., Remillard R.R.,
Nelson R.W., 1998, ApJ 494, L61

\bibitem[]{g93} Greiner J., 1993, IAU Circ. 5786

\bibitem[]{gshkp94} Greiner J., Snowden S., Harmon B.A., Kouveliotou C., 
Paciesas W., 1994, 2nd Compton Symposium, Washington 1993, AIP 304, 260

\bibitem[]{gmr96}  Greiner J., Morgan E.H., Remillard R.A., 1996, ApJ 473, 
L107

\bibitem[6]{gmrph96} Greiner J., Harmon, B.A., Paciesas W.S., Morgan E.H., 
Remillard R.A., 1997, in {\it Accretion Phenomena and Associated Outflows}, 
eds. D.T. Wickramasinghe \etal\, 
ASP Conf. Ser. 121, p. 709

\bibitem[7]{mrmt93} Mirabel I.F., Rodriguez L.F., Marti J.,
             Teyssier R., Paul J., Auriere M.,
              1993, IAU Circ.\ 5773

\bibitem[8]{mr94} Mirabel I.F., Rodriguez L.F., 1994, Nat. 371, 46

\bibitem[8]{mdcr98} Mirabel I.F., Dhawan V., Chaty S., \etal\
1998, AA 330, L9

\bibitem[9]{morg97} Morgan E.H., Remillard R.A., Greiner J., 1997, ApJ 482, 993

\bibitem[10]{niku94} Nagase F., Inoue H., Kotani T, Ueda Y., 1994, IAUC 6094

\bibitem[b]{ob97} Orosz J.A., Bailyn C.D., 1997, ApJ 477, 876

\bibitem[11]{poo96} Pooley G.G., Fender R.P., 1997, MNRAS 292, 925


\end{thebibliography}
\end{document}